%% file: VirtualizationOverhead.tex
\documentclass[conference]{IEEEtran}
\usepackage[T1]{fontenc}
\usepackage{cite}
\usepackage{amsmath,amssymb,amsfonts}
\usepackage{algorithmic}
\usepackage{graphicx}
\usepackage{textcomp}
\usepackage{xcolor}
\usepackage{amsmath,amssymb,amsfonts}
\usepackage{textcomp}
\usepackage{graphicx}
\usepackage{times}
\usepackage{microtype}
\usepackage[UKenglish]{babel}
\usepackage{graphicx,dblfloatfix}
\usepackage{blindtext}
\usepackage{mathtools}
\usepackage{multirow}
\usepackage{tabularx}
\usepackage{subcaption}
\usepackage{hhline}
\usepackage{arydshln}
\usepackage{listings}
\usepackage{adjustbox}
\usepackage{color,soul}
\usepackage[margin=1in]{geometry}
\begin{document}

\title{Evaluating HPC-Style CPU Performance and
Cost in Virtualized Cloud Infrastructures\\}

\author{
    \IEEEauthorblockN{ Jay Tharwani\IEEEauthorrefmark{1},Shobhit Aggarwal\IEEEauthorrefmark{2}, Arnab A Purkayastha\IEEEauthorrefmark{3}}
     \IEEEauthorblockA{\IEEEauthorrefmark{1}Member IEEE, Charlotte, NC, USA\\
    Email: jtharwan@alumni.uncc.edu}
    \IEEEauthorblockA{\IEEEauthorrefmark{2}{Department of ECE}, {The Citadel}, Charleston, NC, USA\\
    Email: shobit.aggarwal@citadel.edu}
    \IEEEauthorblockA{\IEEEauthorrefmark{3}Western New England University, Member IEEE, Springfield, MA, USA\\
    Email: arnab.purkayastha@wne.edu}
    }

\maketitle

\begingroup
\renewcommand\thefootnote{}\footnote{\textbf{This work has been accepted for publication in the proceedings of IEEE Southeastcon 2025. This is the author’s preprint version. The final published version is available on IEEE Xplore.}}
\endgroup

\input{tex/abstract}
\vspace*{0.5 cm}

\input{tex/introduction}
\input{tex/approach}
\input{tex/results}
\input{tex/future}
\input{tex/conclusion}

\bibliographystyle{IEEEtran}
\scriptsize

\bibliography{IJCTT}

\end{document}

%% file: tex/abstract.tex

\begin{abstract}
This paper evaluates HPC-style CPU performance
and cost in virtualized cloud infrastructures using
a subset of OpenMP workloads in the SPEC ACCEL
suite. Four major cloud providers by market share AWS,
Azure, Google Cloud Platform (GCP), and Oracle Cloud
Infrastructure (OCI) are compared across Intel, AMD, and
ARM general purpose instance types under both on-demand
and one-year discounted pricing. AWS consistently delivers
the shortest runtime in all three instance types, yet charges
a premium, especially for on-demand usage. OCI emerges
as the most economical option across all CPU families,
although it generally runs workloads more slowly than
AWS. Azure often exhibits mid-range performance and cost,
while GCP presents a mixed profile: it sees a notable boost
when moving from Intel to AMD. On the other hand, its
ARM instance is more than twice as slow as its own AMD
offering and remains significantly more expensive. AWS’s
internal comparisons reveal that its ARM instance can
outperform its Intel and AMD siblings by up to 49 percent
in runtime. These findings highlight how instance choices
and provider selection can yield substantial variations in
both runtime and price, indicating that workload priorities,
whether raw speed or cost minimization, should guide
decisions on instance types.\end{abstract}

%% file: tex/introduction.tex
\section{Introduction}
\label{sec:Introduction}

High Performance Computing (HPC) often relies on
specialized on-premises clusters to tackle computationally
intensive tasks. As public cloud infrastructures have
matured, many organizations have begun migrating HPC
workloads to providers that offer rapid scalability and
reduced capital expenditures. This shift can simplify
resource provisioning, yet it also introduces questions
about performance variability and cost, particularly when
different processor architectures are available.
Major vendors like Amazon Web Services (AWS),
Microsoft Azure, Google Cloud Platform (GCP), and
Oracle Cloud Infrastructure (OCI) each supply a broad
selection of instance types based on Intel, AMD, and
ARM CPUs. While this variety permits users to tailor
configurations to specific needs, the range of options can
complicate decisions, given that both runtime and pricing
may vary substantially within a single provider’s portfolio.
Evaluating how HPC workloads perform under different
architectures and billing models therefore becomes critical
for achieving optimal throughput without overshooting
budget limits.
The present study focuses on CPU-centric performance
in virtualized cloud environments using a subset of the
SPEC ACCEL 1.2 OpenMP suite. Identical workloads
are executed on Intel, AMD, and ARM-based instances
across all four providers under on-demand and oneyear
discounted pricing. Certain providers demonstrate
faster runtimes but require higher hourly rates, whereas
others emphasize cost savings at the expense of slower
execution. Differences among the three CPU architectures
are also examined, highlighting how each choice can
significantly alter overall runtime and total expenditures.
By documenting these patterns, the analysis aims to guide
readers in balancing performance objectives with cost
considerations when running HPC applications in the
cloud.

%% file: tex/approach.tex
\section{Approach}
\label{sec:method}

\subsection{Market Share-Based CSP Selection}
\begin{figure}[h]
        \centering
                \includegraphics[width=1.0\linewidth, height=0.6\linewidth]{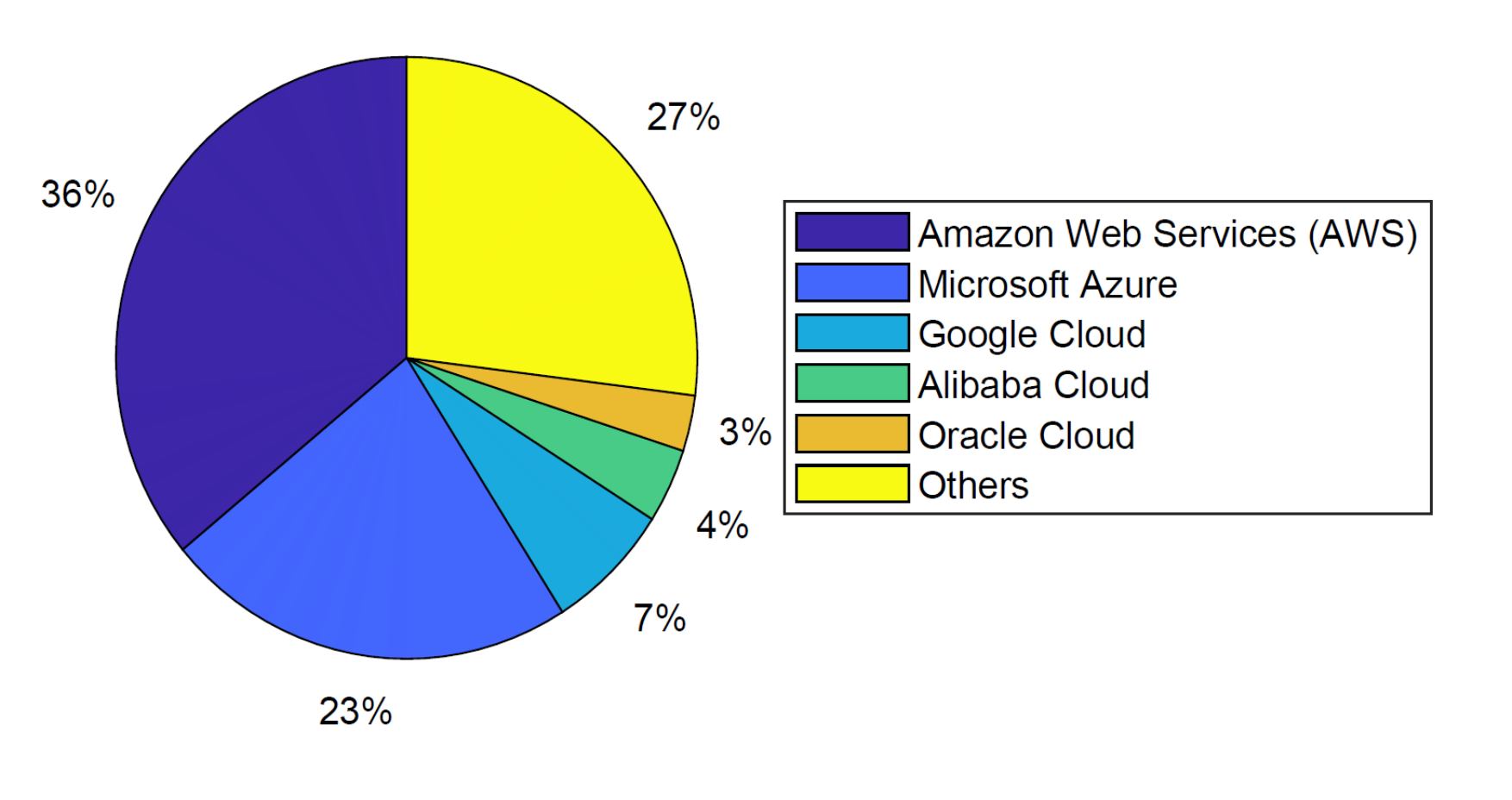}
                \vspace{-5pt}
                \caption{Global Cloud Infrastructure Market Share}
				\label{fig:MarketShare}
\end{figure}

Market share served as the primary criterion for
selecting the cloud service providers used in the instance
comparison.Cloud market share is approximately 300
billion dollars. Amazon Web services leads this market
cap with approximately 36\% stake, followed by Microsoft
Azure with 23\% stake and Google cloud with 7\% stake.
Oracle Cloud Infrastructure was chosen to be the fourth
cloud provider for its quick acceleration towards the
market share. OCI has also recently been involved in
project Stargate as the primary cloud provider for AI
infrastructure. Refer fig \ref{fig:MarketShare}  for a pie chart based representation of the mentioned market share.\cite{stargate}\cite{cloud_market_share}\cite{oracle_fiscal_2025}\cite{oracle_ai_growth}\cite{sgresearch_cloud_growth}

\subsection{Criteria and Rationale for Instance Choice}
Evaluating  efficient options for running general purpose workloads across major cloud providers required
focusing on instance types recognized for popularity and
balanced performance. The selected instances align with
general-purpose requirements for the following reasons:
\begin{itemize}
\item Balanced Resource Allocation: A consistent ratio of compute to memory (e.g., 4 GiB per vCPU)
is maintained, which is well-suited for typical
applications such as web servers, databases, and
application servers.
\item Versatility: Capable of accommodating a wide array of workloads, from development environments to
enterprise-scale applications.
\item Established Popularity: Frequently promoted by
major cloud providers as default general-purpose
option
\item Support for Multiple Architectures: Availability in Intel, AMD, and ARM variants accommodates both traditional and modern computing needs.
\item Cost-Efficiency: Offers reliable performance at lower costs than specialized types, making usage economical for common tasks.
\item Global Availability: Accessible across multiple regions
to ensure consistent pricing and performance,
regardless of location.
All instances from the four cloud providers were
standardized to include 4 virtual processors and 16 GiB of RAM, ensuring a fair basis for performance evaluation.
\end{itemize}

\subsection{Instances Grouped by CPU Type}
This section categorizes the selected cloud instances
according to their hardware platform CPU type. Each
subsection contains instances from multiple providers
that use Intel, AMD, or ARM processors.
\begin{enumerate}
    \item Intel-Based Instances:
    \begin{itemize}
        \item AWS M6i: Custom Intel Xeon Platinum 8375C (Ice
Lake), x86, 4 GiB per vCPU. Use Case: Web servers,
application servers, small-to-medium databases \cite{phoronix_aws_m6i}.
\item Azure Dv5: Intel Xeon Platinum 8370C (Ice Lake)
\cite{azure_dv5_series}, x86, 4 GiB per vCPU. Use Case: Enterprise
applications, relational databases, web services \cite{azure_dv5_dasv5}.
\item GCP N2: Intel Xeon Platinum 8273CL (Cascade
Lake), x86, 4 GiB per vCPU. Use Case: Web and
application servers, enterprise applications, analytics
\cite{gcp_n2_tau_t2a}\cite{gcp_cpu_platforms}.
\item OCI VM.Standard3.Flex: Intel Xeon Platinum 8358
(Ice Lake), x86, 4 GiB per OCPU. Use Case:
Enterprise applications, dynamic web servers, smallto-
medium databases \cite{oci_compute_shapes}.
\end{itemize}
\item AMD-Based Instances:
\begin{itemize}
    \item AWS M6a: AMD EPYC 7R13 (Milan) \cite{azure_dv5_series}, x86, 4
GiB per vCPU. Use Case: Cost-sensitive generalpurpose
workloads, scalable applications.\cite{azure_dasv5_series}
\item Azure Dasv5: AMD EPYC 7763 (Milan), x86, 4 GiB
per vCPU. Use Case: Cost-efficient general-purpose
tasks, database servers \cite{aws_graviton3}.
\item GCP N2D: AMD EPYC 7B12 (Rome), x86, 4
GiB per vCPU. Use Case: Cost-optimized generalpurpose
workloads, scalable applications \cite{gcp_cpu_platforms} \cite{gcp_n2_tau_t2a}.
\item OCI VM.Standard.E4.Flex: AMD EPYC 7742
(Milan), x86, 8 GiB per OCPU. Use Case: Costoptimized
workloads, scalable general-purpose applications
\cite{oci_compute_shapes}.
\end{itemize}
\item ARM-Based Instances:
\begin{itemize}
    \item AWS M7g: AWS Graviton3 [13] (ARM-based,
Neoverse V1 cores), 4 GiB per vCPU. Use Case:
Cloud-native applications, gaming servers, caching
fleets, microservices.
\item Azure Dpsv5: Ampere Altra (80 cores at 3.0 GHz)
[14], ARM, 4 GiB per vCPU. Use Case: Scale-out
workloads, open-source databases, modern cloudnative
applications \cite{azure_dpsv5_dpdsv5}.
\item GCP Tau T2A: Ampere Altra Q64, ARM, 4 GiB
per vCPU. Use Case: Cost-sensitive applications,
containerized microservices, development/test environments\cite{gcp_cpu_platforms} \cite{gcp_n2_tau_t2a}
\item OCI VM.Standard.A1.Flex: Ampere Altra Q80-30,
ARM, configurable memory (up to 64 GiB per
core). Use Case: Cloud-native applications, portable
workloads, web hosting services \cite{oci_compute_shapes}.
\end{itemize}
\end{enumerate}

\subsection{SPEC ACCEL}
SPEC ACCEL is a benchmark suite developed by the
Standard Performance Evaluation Corporation (SPEC)
to evaluate system performance on computationally
intensive parallel applications. It targets parallel pro-gramming models such as OpenCL, OpenACC, and
OpenMP offloading, providing a mix of real-world
workloads and synthetic kernels designed to stress data throughput, parallel execution efficiency, and floating-point computation performance. \cite{specaccel}

The suite includes implementations of widely used
high-performance computing (HPC) kernels and applications, enabling flexible evaluation across heterogeneous hardware and software environments. By focusing on par-allelism and compute intensity, SPEC ACCEL produces results that are representative of workloads in scientific,engineering, and research domains. Version 1.2 of the suite also emphasizes reproducibility, allowing consistent performance comparisons across different platforms and
configurations, making it a widely adopted tool for system characterization.

\subsection{Selected Benchmark}
The SPEC ACCEL suite provides benchmark variants
targeting OpenCL, OpenACC, and OpenMP programming models. In this study, we selected a subset of
the OpenMP benchmarks, with accelerator offloading
explicitly disabled to ensure compilation and execution on the CPU only. This configuration was chosen to isolate CPU performance and measure virtualization overhead without interference from GPU or accelerator-specific optimizations. Benchmark performance was assessed by measuring the total execution time for each selected workload, allowing a comparative analysis across different hypervisors and the bare-metal baseline. The selected
OpenMP benchmarks are summarized below:

\begin{enumerate}
    \item \textbf{pilbdc:} OpenMP version of a kernel derived from an advanced 3-D lattice Boltzmann flow solver, using a two-relaxation-time (TRT-type) collision operator for the D3Q19 model.
    \item \textbf{pseismic:} OpenMP-enabled port of the \texttt{seismic\_PML\_Collino\_3D\_isotropic} solver, a 3D classical split PML program for isotropic media, using a second-order finite-difference spatial operator.
    \item \textbf{psp:} Synthetic CFD benchmark that solves multiple, independent systems of non-diagonally dominant, scalar, pentadiagonal equations in parallel.
    \item \textbf{pep:} Embarrassingly parallel benchmark that generates random number pairs, applies transformations, counts how many fall within specific grid regions (square annuli), and performs a reduction step to aggregate counts and evaluate the pair distribution.
    \item \textbf{pcg:} Conjugate Gradient benchmark that solves an unstructured sparse linear system. It stresses irregular, long-distance communication and relies on unstructured matrix-vector multiplications.
    \item \textbf{pswim:} Weather prediction benchmark using a finite-difference approximation of the shallow-water equations, operating on a 1335$\times$1335 grid over 512 time steps to model fluid dynamics. \cite{specaccel}
\end{enumerate}
    
\subsection{Methodology}
\begin{enumerate}
    \item Instance On-Demand Pricing:

    The pricing information for on-demand instances was gathered from official documentation, pricing tools, and performance reports provided by cloud service providers. Specifications and details for instances were sourced from publicly accessible resources on the AWS, Azure, Google Cloud, and Oracle Cloud Infrastructure websites \cite{aws_pricing,azure_pricing_calculator,gcp_pricing,oci_pricing}.
    
    \item Pricing for One-Year Commitment:

    This study utilized pricing tools and calculators provided by various cloud providers to estimate hourly rates for infrastructure when committed for a one-year term. Cloud providers typically offer reduced pricing for such commitments compared to their on-demand rates.OCI (Oracle Cloud Infrastructure), however, differs from this trend with its universal credits model. Unlike other providers, OCI does not display specific one-year commitment pricing on its website and instead directs users to contact their sales team. Based on available information, an estimated discount of 20\% is commonly applied to their on-demand rates. Due to the lack of precise data, this paper assumes that OCI's on-demand and one-year commitment pricing are equivalent. However, customers may secure a 20\% discount by engaging with the OCI sales team.

    \item SPEC ACCEL Data:

    A subset of the SPEC ACCEL benchmark, as previously described, was executed three times on each of the selected 12 instances. The median runtime from these executions was taken to ensure consistency and repeatability. Each benchmark provides a total runtime metric, which serves as the basis for comparison in this study.
\end{enumerate}

\subsection{Notations and Symbols overview}
The paper uses the following notations and symbols in its data
\begin{itemize}
    \item Cloud: "Cloud" represents the cloud service provider offering the infrastructure on which the instances run. AWS stands for Amazon Web Services, Azure refers to Microsoft Azure, GCP represents Google Cloud Platform, and OCI refers to Oracle Cloud Infrastructure.
    \item Instance Type: Each cloud provider provides a plethora of instance choices . Here , instance type refers to the instance series the cloud service provider is offering. Each instance series is typically characterized by a server type . Each server type is typically characterized by the generation of a particular manufacturers processors \& platform type.
    \item Net(Gbps): Max network bandwidth in Gbps offered by the cloud service provider.
    \item On Demand Pricing:  On-demand pricing billed per hour, when no long term committed contracts have been established.
    \item Discounted Pricing: Estimated per hour pricing when a 1 year contract is in place to use the resources. This is calculated by dividing , the total discounted cost to run the resources per year with the total number of hours in one year.
    \item Runtime: This is the total time(in hrs) elapsed when running the subset of SPEC ACCEL OpenMp workloads , as described earlier , without accelerator offloading.
    \item On Demand Workload Cost: This is the total dollar value the workload will cost with on demand pricing . This is calculated by multiplying the on demand pricing with the total runtime of the workload.
    \item Discounted Workload Cost:This is the total dollar value the workload will cost on the 1 yr discounted pricing .This is calculated by multiplying the per hour one year contract pricing with total runtime of the workload.
\end{itemize}


%% file: tex/results.tex
\section{Outcomes and Insights}
\label{sec:results}

Tables \ref{table:intelInstances} , \ref{table:amdInstances} and \ref{table:armInstances} show the HPC workload runtimes,
on-demand prices, 1 year discounted prices, as well as
how much the HPC workload would cost with on demand
\& 1 year discounted pricing for Intel, AMD and ARM
based instances, respectively.

In the Intel-based instances, AWS completes the workload in about 1.30 hours, while Azure requires around 1.33 hours, OCI 1.32 hours, and GCP 1.62 hours. AWS is therefore the fastest, although its on-demand cost of \$0.25 is nearly 79 percent higher than OCI’s \$0.14 for only about 1.9 percent less runtime. Azure has an on-demand cost of \$0.25, which is also higher than OCI but equal to AWS, and GCP remains the slowest and most expensive in this category at \$0.35. If a one-year commitment is used, the additional expense of AWS compared to OCI shrinks to around 7 percent more, reflecting how longer-term pricing can mitigate AWS’s premium.

In the AMD category, AWS again leads with the
shortest runtime of approximately 0.99 hours. OCI is
slightly slower at 1.05 hours, and both Azure and GCP
come in at about 1.04 hours. Though AWS finishes about
6 percent sooner than OCI, its on-demand total cost of
\$0.17 is more than double OCI’s \$0.078. Azure, at \$0.18
on demand, and GCP, at \$0.20, are also costlier than OCI.
Shifting to one-year pricing narrows AWS’s premium to roughly 67 percent above OCI, so there is still a
substantial cost trade-off for modestly faster performance.

ARM-based instances reveal the largest performance
spread. AWS finishes the workload in about 0.66 hours,
making it substantially faster than the others. OCI’s
1.04 hours is around 36 percent slower than AWS, Azure’s
1.00 hour is about 50 percent slower than AWS, and
GCP’s 2.06 hours represents a more than threefold increase
in runtime. These differences are matched by
equally large cost gaps. AWS’s on-demand cost stands
at \$0.108, whereas OCI’s is just \$0.066. Azure charges
\$0.154, and GCP reaches \$0.32, which is nearly three
times AWS’s on-demand cost and almost five times
OCI’s. Even with a one-year commitment, AWS remains
around 21 percent more expensive than OCI, though
the performance lead is substantial. Azure falls into a
middle ground for both runtime and cost, and GCP is
comparatively expensive, especially when factoring in its
much longer runtime.

\begin{figure}[h]
        \centering
                \includegraphics[width=.85\linewidth, height=0.7\linewidth]{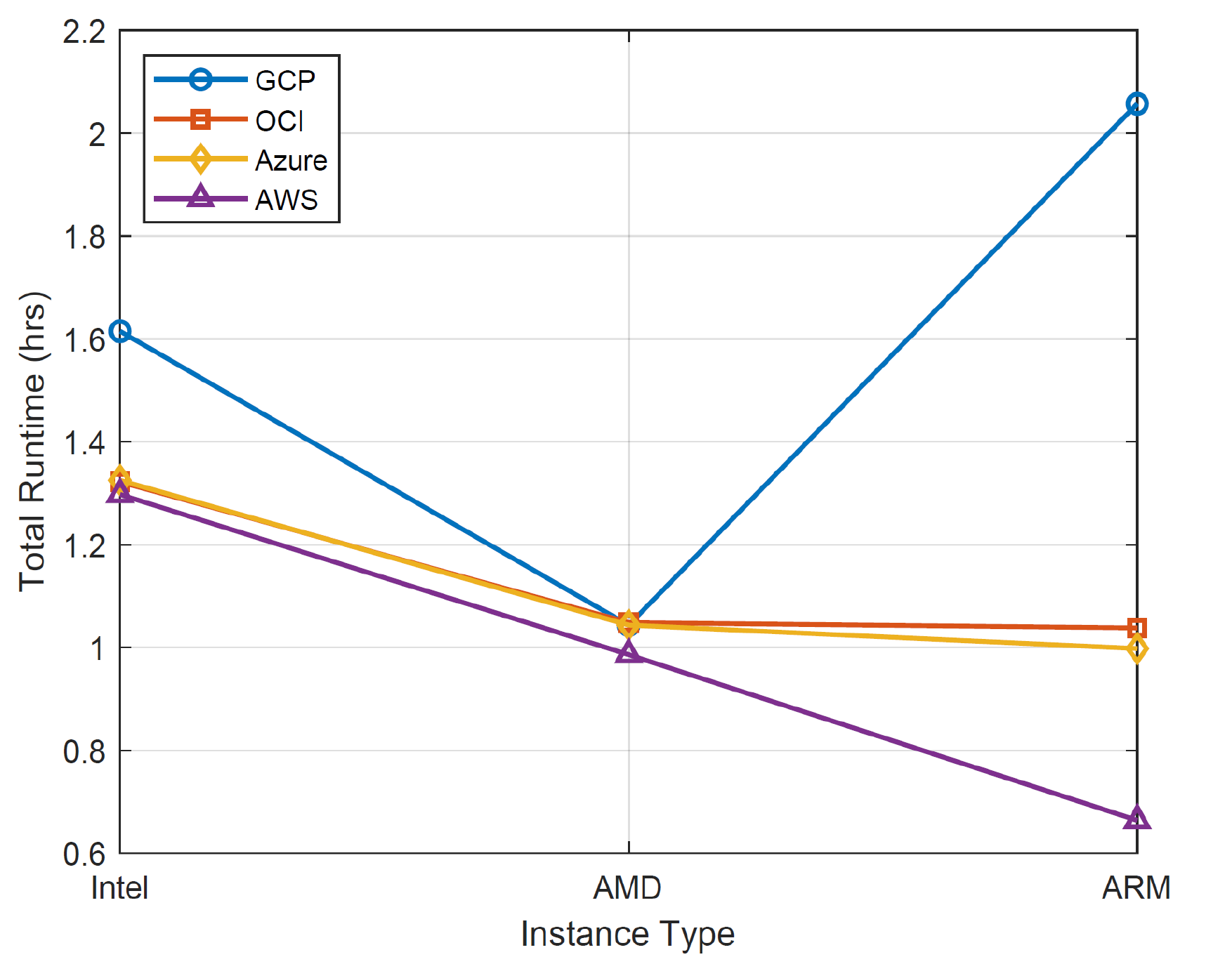}
                \vspace{-5pt}
                \caption{Workload Runtimes across all instances}
				\label{fig:RuntimeAllInstances}
\end{figure}

\begin{figure}[h]
        \centering
                \includegraphics[width=.85\linewidth, height=0.7\linewidth]{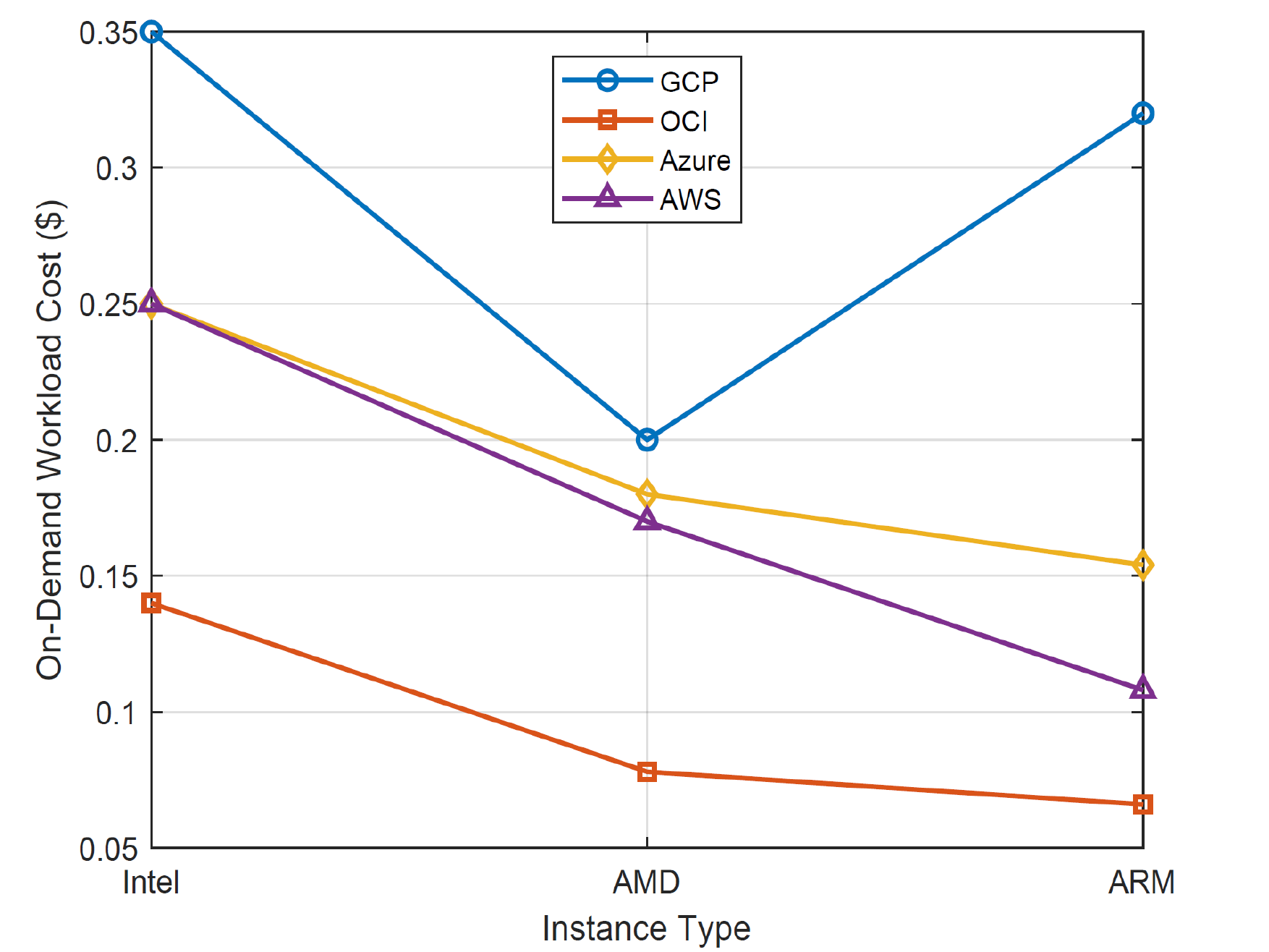}
                \vspace{-5pt}
                \caption{On Demand Workload across all instances}
				\label{fig:OnDemandWorkload}
\end{figure}

\begin{figure}[h]
        \centering
                \includegraphics[width=.85\linewidth, height=0.7\linewidth]{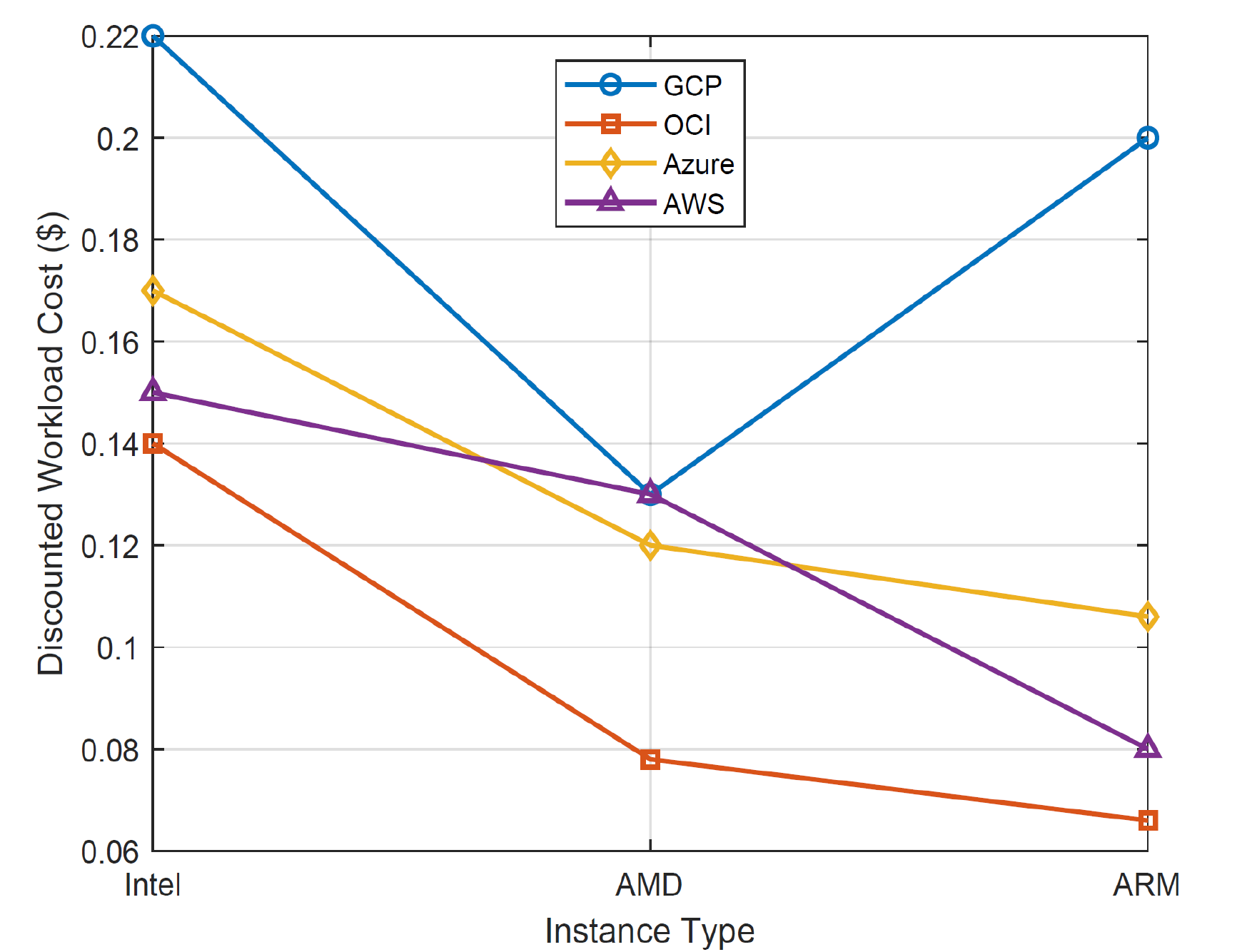}
                \vspace{-5pt}
                \caption{Discounted Workload Cost across all instances}
				\label{fig:DiscountedWorloadCost}
\end{figure}

Figures \ref{fig:RuntimeAllInstances} , \ref{fig:OnDemandWorkload} and \ref{fig:DiscountedWorloadCost} show the workload runtimes ,
on-demand cost and one year discounted costs for all
three instance types and across all clouds .Overall, AWS
consistently offers the quickest runtimes across the three
architectures but at a higher on-demand premium, which
is partially alleviated by a one-year discount. OCI, by
contrast, is the most budget-friendly provider for all
Intel, AMD, and ARM workloads, typically trading off
some performance. Azure often occupies a mid-range
position in both speed and cost, while GCP exhibits
higher runtimes and higher prices, particularly in the
ARM segment where it is over three times slower than
AWS. This analysis underscores that faster completions
generally come at a steeper cost, and one-year discounts
can reduce but do not eliminate that price-to-performance
gap.

When comparing each provider’s offerings across
Intel, AMD, and ARM, notable percentage differences
appear in both runtime and cost. For AWS, the Intel
instance (M6i) at 1.30 hours is about 24 percent slower
than its AMD instance (M6a), which completes the
workload in 0.99 hours, while the ARM variant (M7g)
runs another 33 percent faster than AMD and about
49 percent faster than Intel. AWS’s on-demand cost drops
from \$0.25 on Intel to \$0.17 on AMD (about 32 percent
cheaper) and further down to \$0.108 on ARM (about
57 percent cheaper than Intel). Azure’s Intel instance
(Dv5) at 1.33 hours is roughly 22 percent slower than
its AMD (Dasv5) at 1.04 hours, and that AMD is about
4 percent slower than Azure’s ARM (Dpsv5) at 1.00 hour.
Costs follow a similar pattern, with Azure’s Intel at
\$0.25, AMD at \$0.18 (28 percent cheaper than Intel),
and ARM at \$0.154 (another 14 percent cheaper than
AMD). GCP experiences one of the largest internal gains when moving from Intel (N2) at 1.62 hours to
AMD (N2D) at 1.04 hours, a drop of about 36 percent
in runtime, along with a cost reduction from \$0.35
to \$0.20 (around 43 percent cheaper). However, GCP’s
ARM (Tau T2A) extends runtime to 2.06 hours, roughly
doubling AMD’s duration, and its \$0.32 cost is 60 percent
higher than GCP’s own AMD offering. OCI remains
comparatively balanced: Intel (VM.Standard3.Flex) at
1.32 hours is around 21 percent slower than OCI’s AMD
(VM.Standard.E4.Flex) at 1.05 hours, while both Intel
and AMD are only marginally different from OCI’s ARM
(VM.Standard.A1.Flex) at 1.04 hours. Correspondingly,
OCI’s cost on Intel (\$0.14) is about 44 percent higher
than its AMD rate (\$0.078), and AMD is about 15 percent
more expensive than its ARM cost of \$0.066, underscoring
that OCI stays consistently inexpensive across all
three architectures relative to other clouds.


\begin{table*}[htbp]
\centering
\caption{Intel-Based Instances}
\label{table:intelInstances}
\renewcommand{\arraystretch}{1.6}
\large
\resizebox{0.95\textwidth}{!}{
\begin{tabular}{|p{1.5cm}|p{3.7cm}|p{2cm}|p{2cm}|p{2.3cm}|p{2.3cm}|p{2.5cm}|p{2.5cm}|}
\hline
\textbf{Cloud} & \textbf{Instance Type} & \textbf{Net (Gbps)} & \textbf{Runtime (hrs)} & \textbf{On Demand Pricing} & \textbf{Discounted Pricing} & \textbf{On Demand Workload Cost} & \textbf{Discounted Workload Cost} \\ \hline
GCP & N2 & Up to 10 & 1.61508 & \$0.2187 & \$0.1378 & \$0.35 & \$0.22 \\ \hline
OCI & VM.Standard3.Flex & Up to 10 & 1.32271 & \$0.104 & \$0.104 & \$0.14 & \$0.14 \\ \hline
Azure & Dv5 & Up to 10 & 1.32535 & \$0.192 & \$0.1317 & \$0.25 & \$0.17 \\ \hline
AWS & M6i & Up to 12 & 1.29765 & \$0.192 & \$0.1185 & \$0.25 & \$0.15 \\ \hline
\end{tabular}
}
\end{table*}

\begin{table*}[htbp]
\centering
\caption{AMD-Based Instances}
\label{table:amdInstances}
\renewcommand{\arraystretch}{1.6}
\large
\resizebox{0.95\textwidth}{!}{
\begin{tabular}{|p{1.5cm}|p{3.7cm}|p{2cm}|p{2cm}|p{2.3cm}|p{2.3cm}|p{2.5cm}|p{2.5cm}|}
\hline
\textbf{Cloud} & \textbf{Instance Type} & \textbf{Net (Gbps)} & \textbf{Runtime (hrs)} & \textbf{On Demand Pricing} & \textbf{Discounted Pricing} & \textbf{On Demand Workload Cost} & \textbf{Discounted Workload Cost} \\ \hline
GCP & N2D & Up to 10 & 1.04350 & \$0.19032 & \$0.1199 & \$0.20 & \$0.13 \\ \hline
OCI & VM.Standard.E4.Flex & Up to 10 & 1.04855 & \$0.074 & \$0.074 & \$0.078 & \$0.078 \\ \hline
Azure & Dasv5 & Up to 10 & 1.04358 & \$0.1720 & \$0.1175 & \$0.18 & \$0.12 \\ \hline
AWS & M6a & Up to 12 & 0.98624 & \$0.1728 & \$0.12695 & \$0.17 & \$0.13 \\ \hline
\end{tabular}
}
\end{table*}

\begin{table*}[htbp]
\centering
\caption{ARM-Based Instances}
\label{table:armInstances}
\renewcommand{\arraystretch}{1.6}
\large
\resizebox{0.95\textwidth}{!}{
\begin{tabular}{|p{1.5cm}|p{3.7cm}|p{2cm}|p{2cm}|p{2.3cm}|p{2.3cm}|p{2.5cm}|p{2.5cm}|}
\hline
\textbf{Cloud} & \textbf{Instance Type} & \textbf{Net (Gbps)} & \textbf{Runtime (hrs)} & \textbf{On Demand Pricing} & \textbf{Discounted Pricing} & \textbf{On Demand Workload Cost} & \textbf{Discounted Workload Cost} \\ \hline
GCP & Tau T2A & Up to 10 & 2.05722 & \$0.154 & \$0.09702 & \$0.32 & \$0.20 \\ \hline
OCI & VM.Standard.A1.Flex & Up to 10 & 1.03820 & \$0.064 & \$0.064 & \$0.066 & \$0.066 \\ \hline
Azure & Dpsv5 & Up to 10 & 0.99804 & \$0.1540 & \$0.1058 & \$0.154 & \$0.106 \\ \hline
AWS & M7g & Up to 12 & 0.66395 & \$0.1632 & \$0.1199 & \$0.108 & \$0.080 \\ \hline
\end{tabular}
}
\end{table*}

\vspace{-10pt}

\vspace{-10pt}

\vspace{-10pt}

%% file: tex/future.tex
\section{Future}
\label{sec:future}

The present study focuses on single-node instances
running a subset of the SPEC ACCEL 1.2 OpenMP suite.
In the past a similar , non-HPC study has been done using
Geekbench \cite{tharwani2024}. Future research could address several
additional dimensions to broaden the understanding of
HPC workloads in cloud environments. One direction
involves examining multi-node or cluster configurations
to assess how providers’ high-speed networking and
interconnects influence scalability and performance.
Future efforts might also investigate specialized HPC

Future efforts might also investigate specialized HPC
instance families that offer features such as remote direct
memory access (RDMA) or local NVMe storage to see
whether they can further reduce runtime or improve
cost efficiency. Similarly, performing experiments under
containerized frameworks (for instance, Kubernetes or
Docker Swarm) could reveal how orchestration overhead
impacts HPC workloads on various architectures. Finally,
the interplay between different pricing strategies—beyond
one-year discounts—such as spot instances or reserved
instances over multi-year terms, deserves further examination for organizations seeking to optimize cloud spending over longer horizons. Finally also from a cost standpoint, this study only takes into account the pricing of the vCPUs \& memory in its pricing data . Future studies can take into account the storage variability into its pricing as well.

%% file: tex/conclusion.tex
\section{Conclusion}
\label{sec:Conclusion}

This paper presents an evaluation of HPC-style CPU
performance and cost in four major cloud environments—
Amazon Web Services, Microsoft Azure, Google
Cloud Platform, and Oracle Cloud Infrastructure—using
Intel, AMD, and ARM instance families. The results,
obtained from a subset of the SPEC ACCEL 1.2 OpenMP
benchmarks, reveal consistent trade-offs between runtime
and expense. AWS generally delivers the fastest completion
times but at a higher on-demand cost, while
OCI emerges as a lower-cost option with modestly
slower runtimes. Azure and GCP occupy intermediate
positions, though GCP’s ARM offering in particular lags
significantly. These patterns largely persist under one-year
discounted plans, albeit with reduced pricing differentials.

The findings underscore the importance of aligning
HPC workloads with appropriate architectures and pricing
models. While high speed can be purchased at a premium,
slower or more cost-effective alternatives may suffice for
less time-sensitive workloads. Since cloud providers continue
to refine their hardware offerings and adjust pricing
strategies, the optimal choice for HPC workloads will
likely evolve. The analyses provided here serve as a guide
for practitioners seeking to balance performance and cost,
and they highlight avenues for future investigation around
distributed computing, specialized HPC instances, and
emerging pricing schemes.